\documentclass[10pt,a4]{article}
\usepackage[cdot,squaren, thinspace,thinqspace]{SIunits}
\usepackage{braket}
\usepackage{dsfont}
\usepackage{graphicx}
\usepackage{supertab}
\usepackage{booktabs}
\usepackage{cite}
\usepackage{times}

\newcommand{\subs}[2]{\hspace{-1.1mm}  {}_{ #1}  \hspace{0.7mm} {}_{#2} \hspace{-1.1mm}}

\newcommand{\subr}[1]{\hspace{-1.1mm} {}_{ #1}}

\begin{document}


\title{Producing high fidelity single photons with optimal brightness via waveguided parametric down-conversion}
\date{}
\author{ K.~Laiho,  K.~N.~Cassemiro and Ch.~Silberhorn \\ \small{Max Planck Institute for the Science of Light} \\ \small{G\"unther-Scharowsky-Strasse 1/Building 24, D-91058 Erlangen, Germany} \\ \small{Kaisa.Laiho@mpl.mpg.de}}\maketitle


\maketitle


\begin{abstract}
Parametric down-conversion (PDC) offers the possibility to control the fabrication of non-Gaussian states such as Fock states. However,  in conventional PDC sources energy and momentum conservation introduce strict frequency and photon number correlations, which impact the fidelity of the prepared state. In our work we optimize the preparation of single-photon Fock states from the emission of waveguided PDC via spectral filtering. We study the effect of correlations via photon number resolving detection and quantum interference.  Our measurements show how  the reduction of mixedness due to filtering can be evaluated. Interfering the prepared photon with a coherent state we establish an experimentally measured fidelity of the produced target state  of  78\%. 
\end{abstract}



\section{Introduction}

A practical single photon source should meet the demands of high fidelity, brightness and efficiency. As a source of photon pairs, parametric down-conversion (PDC) enables conditional state preparation---heralding on one of the twin photons  (idler) gives exact timing information about the other (signal). In addition to this, single pass PDC sources have the advantage of operational simplicity and the quality of the quantum state preparation is mainly limited by the available detectors.

Recent experiments have demonstrated a full tomographic characterization of  Fock states or photon added/subtracted states prepared from  the PDC emission in the low gain regime~\cite{A.I.Lvovsky2001,A.Zavatta2004,A.Ourjoumtsev2006,Ourjoumtsev2007,V.Parigi2007}. Nevertheless, since the photon-number distribution of signal and idler is thermal,  higher photon-number components degrade the fidelity of the prepared target state. This effect becomes prominent at higher gains i.e. if high preparation rates are needed. In addition to photon-number correlation, spectral correlation between the  PDC photon pairs cannot be avoided in conventional sources~\cite{Y.-H.Kim2004,W.Wasilewski2006,Poh2007,J.Chen2009}.  Although the spectral correlation is a useful resource for the generation of  highly entangled states~\cite{M.Hendrych2009, M.V.Fedorov2007,  M.Avenhaus2008}, the preparation of non-Gaussian states in one of the twin modes leads to mixedness whenever the other field mode is traced out ~\cite{A.B.U'ren2005}.

In the past, the manipulation of the spectral properties has  been in the interest of  broadband and efficient  non-linear classical frequency conversion~\cite{Diels2006}. More recently, tailoring these properties has been suggested as a method of source engineering for quantum optics~\cite{T.E.Keller1997, W.P.Grice2001}. Several sophisticated methods have been proposed for state decorrelation, which unfortunately often require means that are available only under specific conditions~\cite{P.J.Mosley2008,  Valencia2007, M.Avenhaus2009, O.Cohen2009,M.Halder2009}. Nevertheless, the decorrelation of the PDC photons is always possible via spectral filtering, which drastically changes the shape of the spectral correlation function. The influence of the spectral correlation on the  purity of a single photon state prepared by filtering was first experimentally demonstrated  by Wasilewsky {\it et al.}~\cite{W.Wasilewski2007}.

In addition to the optimal state preparation, a precise  characterization of the prepared photonic states both in photon-number and spectral degrees of freedom is required.
The latest developments in photon counting technologies allows to characterize photonic states via the direct detection of the photon-number statistics. One of these techniques applicable for pulsed light is time-multiplexed detection (TMD), which enables a loss-tolerant reconstruction of photon statistics~\cite{D.Achilles2005, Avenhaus2008}.  
Regarding the spectral properties, their impact on the state preparation can be studied via Hong-Ou-Mandel (HOM) type interference, originally applied to observe the temporal length of PDC photons~\cite{C.K.Hong1987}. This experiment can be implemented with a symmetric beam splitter  and a coincidence counting device. Two indistinguishable photons impinging simultaneously on the different ports of the beam splitter  bunch together and no coincidences are detected. Thus, the measurement of HOM interference dip between signal and idler is a measure of their spectral indistinguishability, or in other words  of their equity~\cite{M.Avenhaus2008}. 
However, for characterizing the fidelity of the prepared single photon, one must employ an independent, spectrally pure reference field,  which can be a highly attenuated coherent state. The measurement of HOM interference between these two independent sources allows to quantify their spectral overlap, as evidenced in several studies~\cite{J.G.Rarity1997, T.B.Pittman2003, T.B.Pittman2004, J.G.Rarity2005,Kolenderski2008}. Considering the direct probing of more evolved non-Gaussian states, such as displaced single photons, the determination of this overlap is essential \cite{K.Laiho2009}.

Lately, PDC sources with extreme high brightness have been accomplished by employing waveguided structures rather than bulk crystals~\cite{M.Fiorentino2007}. As filtering happens at the cost of photon flux, waveguides (WGs) can offer benefits for preparing high fidelity single photons. In this paper we  show how to optimally adapt the bandwidth of the filter such that the trade-off between the fidelity and brightness of the source is balanced. In Sec.~\ref{sec_ellipse} we provide easy applicable tools for analyzing and estimating the spectral correlation of a waveguided PDC source and give a recipe of how to design a filter for optimal state decorrelation.
The spectral properties and the performance of the employed filters are verified in Sec.~\ref{sec_decorrelation}. 
Our main focus, the characteristics of the filtered state is presented in Sections \ref{sec_statistics}  and  \ref{sec_HOM}; we study the statistics of the heralded state and finally determine the spectral overlap between signal and reference. This allows us, to our knowledge for the first time,  to directly extract  
the overall fidelity of a heralded single photon by utilizing the HOM interference.

\section{\label{sec_ellipse} Properties of the spectral correlation function}
The quantum state of photon pairs generated via PDC  is usually described as
\begin{equation} 
\ket{\psi} =\ket{0}+\zeta \int \int d\omega_{s}d \omega_{i}\phi(\omega_{s}, \omega_{i})\ket{\omega_{s},\omega_{i}} + O(\zeta^{2}) \;,
\label{eq_pdc_lowgain}
\end{equation}
where $|\zeta|^{2}\ll 1$ is the pair creation probability proportional to the pump power, and the labels $s,i$ refer to signal and idler. The spectral correlation function $\phi(\omega_{s}, \omega_{i})$ arises as a consequence of energy and momentum conservation. The former is constrained by the spectrum of the pump pulse $\alpha(\omega_{s},\omega_{i})$ and the latter, being governed by the dispersion of the non-linear optical medium of the length $L$,  is characterized by 
the phase-matching (PM) function $\Phi(\omega_{s},\omega_{i}) = \textrm{sinc}(\Delta k(\omega_{s},\omega_{i})L/2)e^{i \Delta k(\omega_{s},\omega_{i})L/2}$.
We study the spectral correlation in the vicinity of the central frequencies of pump, signal and idler $\omega^{0}_{\mu}$ ($\mu =p, s, i$), 
at which $\hbar \omega_{p}^{0} = \hbar \omega_{s}^{0}+\hbar \omega_{i}^{0}$ and $\Delta k^{0} = k_{s}(\omega^{0}_{s})+k_{i}(\omega^{0}_{i})-k_{p}(\omega^{0}_{p}) - \frac{2\pi}{\Lambda} = 0$.  The grating period $\Lambda$  of the quasi-phasematched structure can be used for adapting the desired central wavelengths. (See e.g.~\cite{A.B.U'ren2005}).

\begin{figure}[!htb]
\begin{center}
\includegraphics[width = 0.62\textwidth]{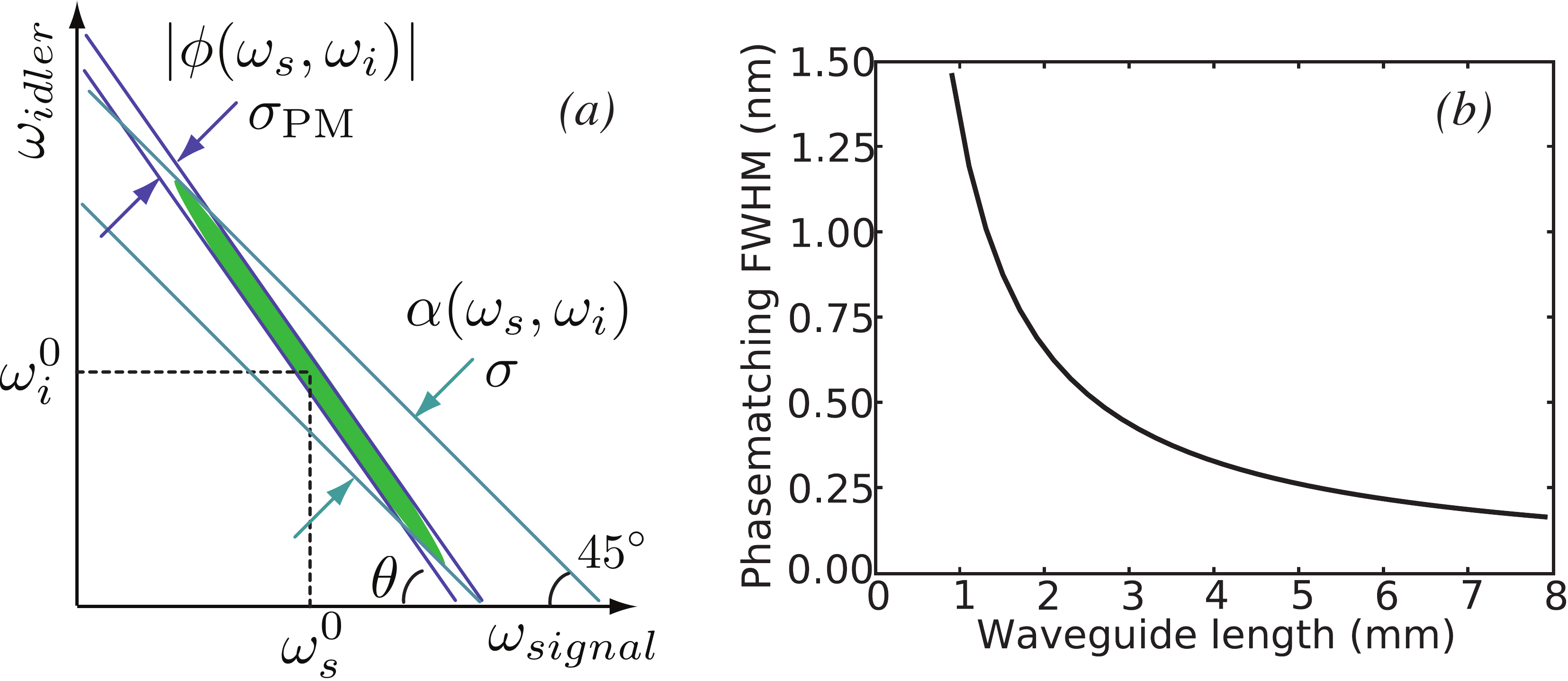}
\caption{(a) Illustration of the spectral correlation following from the pump and PM properties. (b) Predicted FWHM (full width at half maximum) of the PM  at \unit{796}{\nano\meter} with respect to WG length. Employed WG  parameters:   \unit{4x6}{\micro \meter^{2}} cross section and \unit{7.46}{\nano \meter} poling period. }
\label{fig_correlation}
\end{center}
\end{figure}

For simplicity, we describe the spectrum of the pump by a Gaussian envelope
 \begin{equation}
\alpha(\nu_{s},\nu_{i}) = e^{-\frac{1}{\sigma^{2}}(\nu_{s} + \nu_{i} )^{2}},
\label{eq_pump}
\end{equation} 
where  $\nu_{\mu}=\omega_{\mu}-\omega^{0}_{\mu}$ is the detuning from the central wavelength. In addition, we expand the phase-mismatch $ \Delta k(\omega_{s},\omega_{i})$ appearing in the PM function in first order Taylor series around $\omega^{0}_{\mu}$ and use the Gaussian approximation for the sinc-function. Thus, the PM function can be estimated by
\begin{equation}
\Phi(\nu_{s},\nu_{i}) \approx e^{-\frac{\gamma L^{2}}{4}(\kappa_{s}\nu_{s} +  \kappa_{i}\nu_{i} )^{2}} e^{i\frac{L}{2}(\kappa_{s}\nu_{s} +  \kappa_{i}\nu_{i})},
\label{eq_PM}
\end{equation}
where $\kappa_{\mu} = k^{\prime}_{\mu}(\omega^{0}_{\mu})- k^{\prime}_{p}(\omega^{0}_{p}) $ is determined by the group-velocity mismatch of the non-linear medium and the parameter $\gamma = 0.193$ adapts the width of the Gaussian to the width of the original sinc-function. Finally, the amplitude of the spectral correlation function is written as
\begin{eqnarray}
\left | \phi(\nu_{s}, \nu_{i}) \right | &=& \alpha(\nu_{s},\nu_{i}) \left |\Phi(\nu_{s},\nu_{i})  \right | \nonumber \\
&&\hspace{-75pt}\approx \exp   \Bigg( {-\left[ \begin{array}{cc} \nu_{s}&\nu_{i}  \end{array} \right]  
\underbrace{\left[\begin{array}{cc}  
\frac{1}{\sigma^{2}}+ \gamma \frac{L^{2}}{4}\kappa_{s}^{2} & 
\frac{1}{\sigma^{2}}+ \gamma \frac{L^{2}}{4}\kappa_{s}\kappa_{i} \\
 \frac{1}{\sigma^{2}}+ \gamma \frac{L^{2}}{4}\kappa_{s}\kappa_{i} & 
 \frac{1}{\sigma^{2}}+ \gamma \frac{L^{2}}{4}\kappa_{i}^{2} \end{array} \right]}_{\mathcal{M}} 
 \left[\begin{array}{c} \nu_{s}\\ \nu_{i}  \end{array} \right] } \Bigg ) ,
\label{eq_ellipse}
\end{eqnarray}
which represents an ellipse in the $(\nu_{s},\nu_{i})$--space as illustrated in Fig.~\ref{fig_correlation}(a). For a decorrelated and separable photonic state the ellipse is oriented parallel to the $(\nu_{s},\nu_{i})$--coordinates, i.e. the matrix $\mathcal{M}$ in Eq.~(\ref{eq_ellipse}) is diagonal.  Non-vanishing off-diagonal terms correspond to a correlated photonic state. We stress that these properties can be tailored by modifying the pump bandwidth or the PM properties, e.g. by adjusting the waveguide length.

The width of the ellipse axes are determined by the eigenvalues of  $\mathcal{M}$ and are equal to
\begin{equation}
\frac{1}{\sigma^{2}_{\textrm{ellipse axes}}} = \frac{1}{\sigma^{2}}+ \gamma(\kappa_{s}^{2}+ \kappa_{i}^{2}) \frac{L^{2}}{8} \pm \sqrt{ \frac{1}{\sigma^{4}} + \gamma^{2}(\kappa_{s}^{2}+ \kappa_{i}^{2})^{2}\frac{L^{4}}{64} +\gamma\frac{\kappa_{s}\kappa_{i}}{\sigma^{2}} \frac{L^{2}}{2}}\;,
\label{eq_axes}
\end{equation}
whereas the tilt can be extracted from the eigenvectors. In the limit of  spectrally  broadband pump ($1/\sigma^{2} \approx 0$) the tilt $\theta$  is given by  $\tan(\theta) = \kappa_{s}/\kappa_{i}$ and the minor axis of the ellipse is purely determined by the PM properties,
\begin{equation}
\frac{1}{\sigma_{\textrm{PM}}^{2}} = \gamma (\kappa_{s}^{2}+ \kappa_{i}^{2})\frac{L^{2}}{4}\;.
\end{equation}
In Fig.~\ref{fig_correlation}(b) we show the behavior of this width as a function of $L$ considering the expected material properties of our WG. Even though this theoretical prediction gives an indication about the PM properties, the actual spectral correlation function needs to be accurately  characterized. This is especially important for waveguided PDC as the non-ideal properties of the guide typically influence the PM function significantly.

 In terms of Fock state preparation, the joint amplitude of signal and idler in Eq.~(\ref{eq_ellipse}) has to be decoupled. To optimally achieve this goal we employ a short WG, thus broadening the PM bandwidth, and a femtosecond pump, which allows the state decorrelation via a proper design of the spectral filter. Filtering reduces the width of the ellipse's major axis and modifies the correlation such that 
 the ellipse evolves into a circular shape or has a very small remaining tilt. Thence, the bandwidth of the filter has to be matched with the width of the ellipse's minor axis.

\subsection{Simple estimation of the ellipse's angle and axis}
One of the most direct ways to characterize the tilt of the spectral correlation ellipse is to measure the unconditional spectral marginal distributions of signal and idler. The tilt is determined by the simple relation $\tan(\theta) = \frac{\Delta \omega_{i}} {\Delta \omega_{s}}$, where $\Delta \omega_{\mu} \ (\mu = i,s)$ is the FWHM of the marginal. However, several spatial modes can propagate in the WG and an accurate determination of the tilt via this method becomes difficult~\cite{A.Christ2009}. Another possibility relies on detuning the pump  wavelength and observing the respective change in the central wavelengths of signal and idler. Similarly, we can also employ narrowband filters in signal and idler arms, and record the wavelengths at which photon-number correlation appears, thus mapping the spectral correlation function.  

\begin{figure}[!htb]
\begin{center}
\includegraphics[width = 0.9\textwidth]{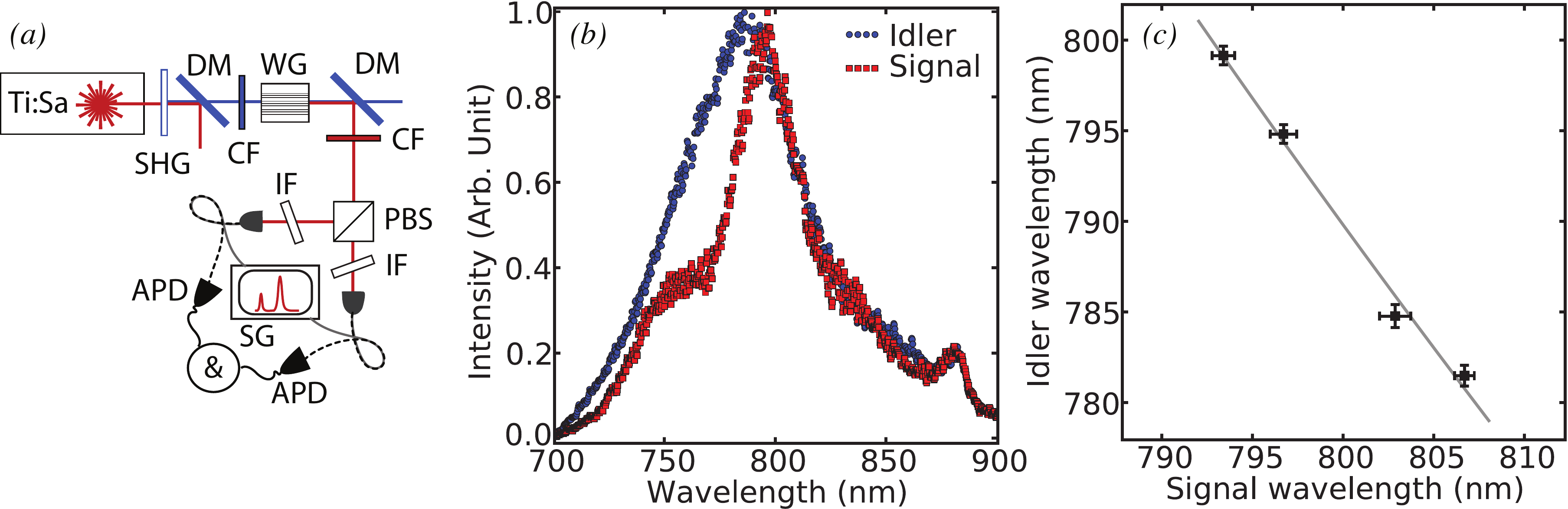}
\caption{(a) Setup for measuring the tilt of the PM function. [The list of abbreviations appears in Fig.~\ref{fig_setups}.] (b) Nearly degenerate, measured marginal distributions of signal and idler and (c)  the wavelengths at which coincidence counts between signal and idler were recorded.  }
\label{fig_marginals}
\end{center}
\end{figure}

In all PDC measurements presented in this paper we  employed a \unit{2.1}{\milli \meter} long, type-II, periodically poled KTP WG, which was pumped by a frequency doubled Ti:Sapphire laser operating at ultrafast regime (\unit{796}{\nano \meter}, \unit{10}{\nano \meter} FWHM, \unit{170}{\femto \second} autocorrelation length and \unit{1-4}{ \mega \hertz} repetition rate). 
For the estimation of the ellipse's tilt we separated signal and idler photons in a polarizing beam splitter and launched  them into single mode (SM) fibers. The single photon spectra was measured with a sensitive spectrograph (Andor) and the photon-number correlation with avalanche photo diodes (APDs) as illustrated in  Fig.~\ref{fig_marginals}(a). First of all, we measured the spectral marginal distributions of signal and idler as shown in Fig.~\ref{fig_marginals}(b). After that, two narrowband interference filters were set on the beam paths of signal and idler and the photon-number correlation was mapped by tuning the filters. For the accurate determination of the ellipse's tilt the filter wavelengths were recorded with the spectrograph. From the results shown in Fig.~\ref{fig_marginals}(c) we extracted the following value for the tilt: $\theta = 54.7^{\circ}\pm 1.5^{\circ}$.  

Regarding the major axis of the correlation ellipse, note that its 
width is highly dependent on the bandwidth of the pump as illustrated in  Fig.~\ref{fig_correlation}(a). The FWHM of the major axis  can be roughly estimated  from the marginal distributions by $\Delta \omega_{s} / \cos(\theta) \approx \Delta \omega_{i} / \sin(\theta)$. Our results indicate that pumping the WG with  \unit{1-1.5}{\nano\meter} broad pulse leads to an approximately \unit{35-40}{\nano\meter} broad major axis, which increases to  values between \unit{55-60}{\nano\meter}, if the pump is  \unit{2-2.5}{\nano\meter} broad.

\begin{figure}[!htb]
\begin{center}
\includegraphics[width = 0.8\textwidth]{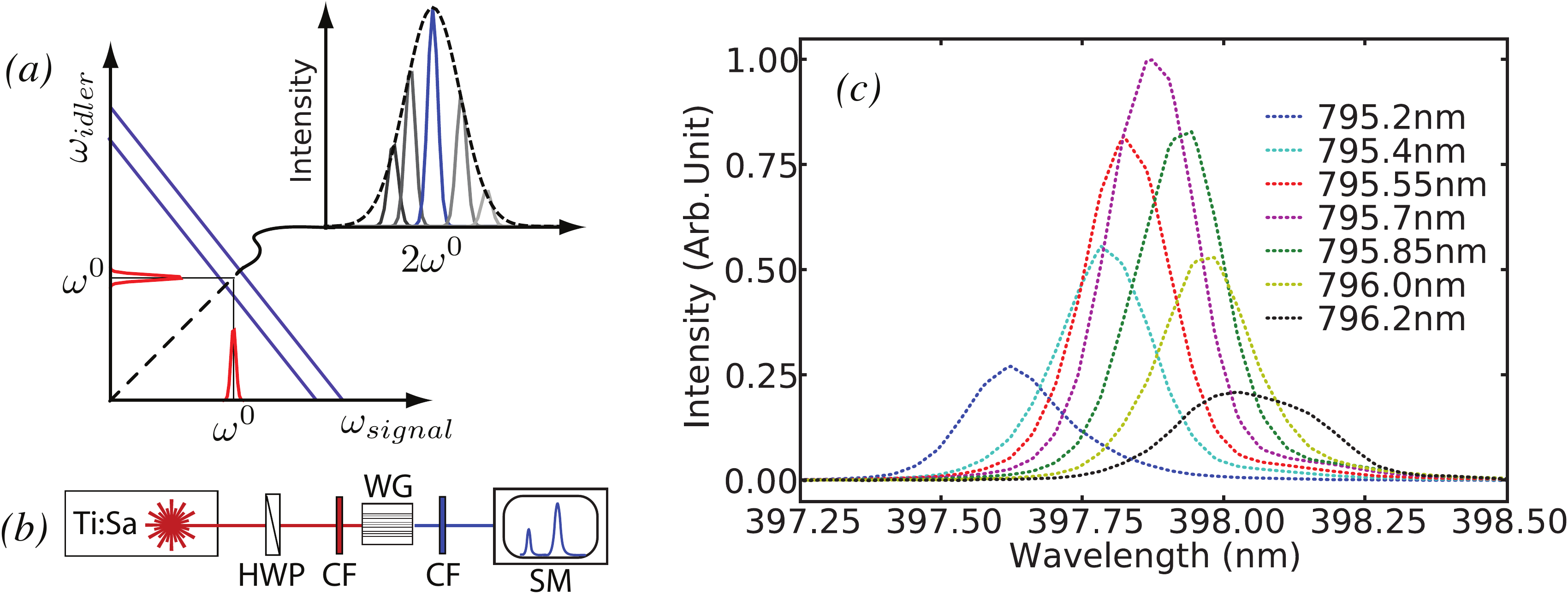}
\caption{(a)  Illustration of how to scan the PM structure with SH measurements. (b) Setup employed in the SH measurement. [The list of abbreviations appears in Fig.~\ref{fig_setups}.] (c) Measured SH responses to different fundamental wavelengths.}
\label{fig_shg_response}
\end{center}
\end{figure}

According to Fig.~\ref{fig_correlation}(b) the exact determination of the PM bandwidth of our WG, i.e. the minor axis of the ellipse, requires sub-nanometer resolution. Instead of employing sensitive monochromators, we studied the second harmonic (SH) response of the WG. In this process two photons at the fundamental frequency give rise to a single up-converted one. Therefore,  
the fundamental probe always lies on the line $\omega_i=\omega_s$ of the $(\omega_{s},\omega_{i})$--space, as illustrated in Fig.~\ref{fig_shg_response}(a).  In other words, tuning the frequency of the fundamental probe and measuring the respective SH response allows us to characterize the width of the PM function. If the fundamental probe is narrow with respect to the PM bandwidth $\sigma_{PM}$, and the tilt of the PM only slightly deviates from 45$^{\circ}$, the envelope of the SH response is approximately equal to
\begin{equation}
I( \nu_{\mathrm{SH}}) \sim \exp{\left( -\frac{\nu_{\mathrm{SH}}^{2}}{\sigma_{\mathrm{PM}}^{2}} \right)}\; .
\end{equation}

We recorded the SH response with a grating spectrometer by tuning the fundamental probe (\unit{795.7}{\nano\meter}, \unit{0.6}{\nano\meter} FWHM, and \unit{2.6}{\pico\second} autocorrelation length) in steps of  approximately \unit{0.2}{\nano\meter} over the phase matching curve [setup shown in Fig.~\ref{fig_shg_response}(b)]. In our case, the measurement is limited by both the resolution of the spectrograph (\unit{0.36}{\nano\meter} at \unit{800}{\nano\meter}) and the bandwidth of the fundamental probe beam. Nevertheless, our results shown in Fig.~\ref{fig_shg_response}(c) indicate that we can probe the shape of the PM function and that its bandwidth is on the order of the probe bandwidth.

\begin{figure}[!htb]
\begin{center}
\includegraphics[width = 0.8\textwidth]{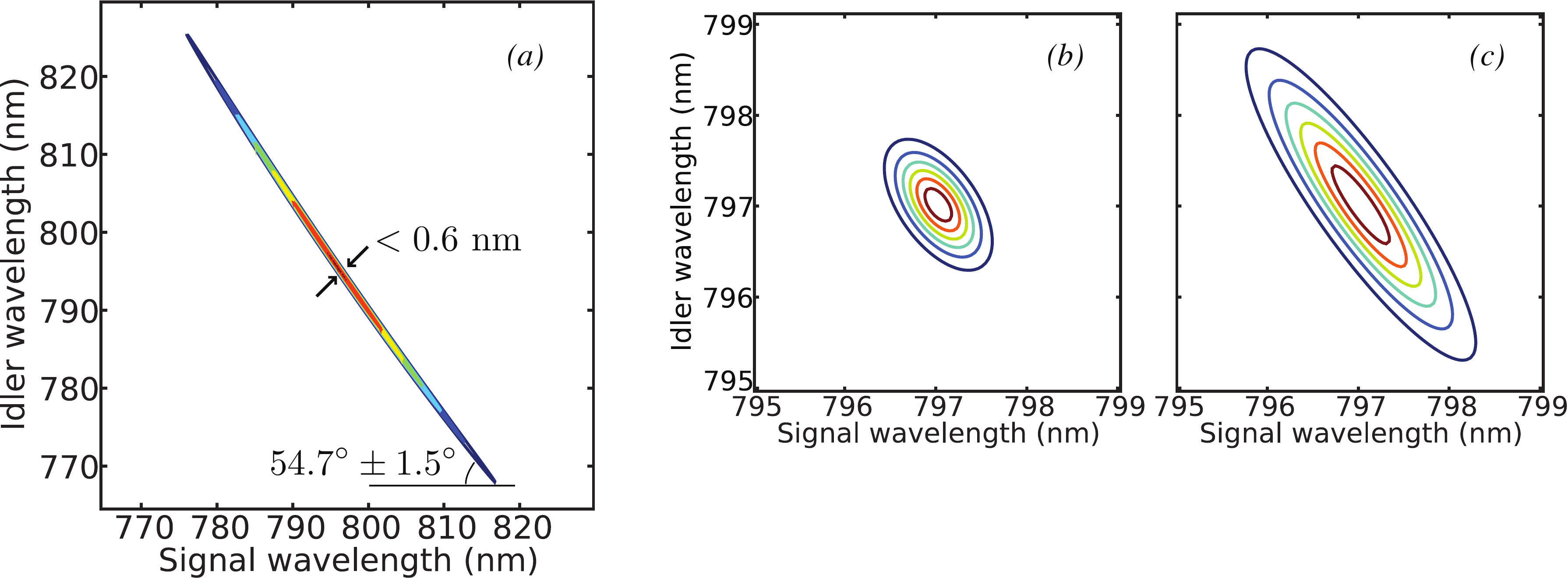}
\caption{(a) The reconstructed spectral correlation ellipse $|\phi(\nu_{s}, \nu_{i})|^{2}$. We applied in the simulation \unit{2.5}{\nano\meter}  broad pump at centered at \unit{795.7}{\nano\meter}. Correlations remaining between signal and idler after (b) \unit{1}{\nano\meter} and (c) \unit{2.5}{\nano\meter}  broad spectral filtering.}
\label{fig_ellipse_rec}
\end{center}
\end{figure}

By combining our findings we are able to reconstruct the correlation ellipse $|\phi(\nu_{s}, \nu_{i})|^{2}$ of our photon source. Using the estimated values of  the tilt and width of the PM,  and taking into account a pump with FWHM of \unit{2.5}{\nano\meter} we obtain the ellipse shown in Fig.~\ref{fig_ellipse_rec}(a). The result implies that signal and idler are highly frequency correlated, which is a consequence of the long interaction length of the PDC photons in the WG. Further, our result suggests that the optimal filter bandwidth is on the order of \unit{0.6}{\nano \meter}. Unfortunately, spectral filters at \unit{800}{\nano \meter} with high transmission and narrow bandwidth are rarely available. In consequence,
 we study in the next section the decorrelation of the generated photonic state by employing spectral filters with \unit{1}{\nano\meter} and \unit{2.5}{\nano\meter} bandwidths. We expect slight remaining correlations after sending the prepared state through these filters as shown in Figs~\ref{fig_ellipse_rec}(b) and \ref{fig_ellipse_rec}(c), respectively. 
 
\section{\label{sec_decorrelation}Decorrelation of signal and idler}
The indistinguishability between signal and idler can be studied via a HOM interference, where signal and idler are superimposed in a 50/50 beam splitter. The full visibility is achieved either if the correlation function is circular, or if it is symmetric with respect to the axis of energy preservation, i.e. if the correlation ellipse is oriented at $45^{\circ}$~\cite{W.P.Grice1997}. Here we employ this measurement to test and corroborate the acquired knowledge about the spectral ellipse.

We adapt the analysis of Avenhaus et al.~\cite{M.Avenhaus2008} to the study of the HOM visibility with spectral filter. The interference between signal and idler can be characterized by the overlap factor
\begin{equation}
\mathcal{O} = \frac{\int \int  d\nu_{s} d\nu_{i} \phi(\nu_{s}, \nu_{i}) \phi^{*}(\nu_{i}, \nu_{s})}{\int \int  d\nu_{s} d\nu_{i}\left| \phi(\nu_{s}, \nu_{i})\right |^{2}}\;.
\label{eq_O}
\end{equation}
Using the Gaussian approximation of the spectral correlation function, the overlap factor can be expressed as a product of a spectral ($s$) and temporal ($t$) overlap i.e., $\mathcal{O} =\mathcal{O}_{s} \mathcal{O}_{t} $.
While spectral mismatch results from an asymmetric spectral correlation function, temporal mismatch originates from the phase terms, which are ultimately related to the properties of the birefringent media. These can be inferred from the properties of the correlation ellipse via Eq.~(\ref{eq_PM}). If perfect overlap was necessary, the temporal mismatch could be compensated with an optical delay line.

The defined overlap factors depend on the ellipse's tilt $\theta$  and aspect ratio $\mathcal{A}$, defined as the ratio between the major and minor axes. They can be expressed as
\begin{equation}
\mathcal{O}_{s} = \frac{2\,\mathcal{A}}{\sqrt{(1+\mathcal{A}^{4})(1-\sin^{2}{(2\theta)})+ 2\,\mathcal{A}^{2}(1+\sin^{2}{(2\theta)})}} \quad \textrm{and}
\label{eq_O_spectral}
\end{equation}
\begin{equation}
\mathcal{O}_{t} = \exp{\left({-\frac{\mathcal{A}^{2}}{2\gamma} \; \frac{ [1-\sin^{2}({2\theta})]\,\mathcal{A}^{2} + [1-\sin({2\theta})]^{2} }{ [1+\mathcal{A}^{4}][1-\sin^{2}({2\theta})]+ 2\,\mathcal{A}^{2}[1+\sin^{2}({2\theta})] }}\right)}\;,
\label{eq_O_temporal}
\end{equation}
where we take use of an estimation that the PM width and the ellipse's minor axis  are equal.

\begin{figure}[!htb]
\begin{center}
\includegraphics[width = 0.77\textwidth]{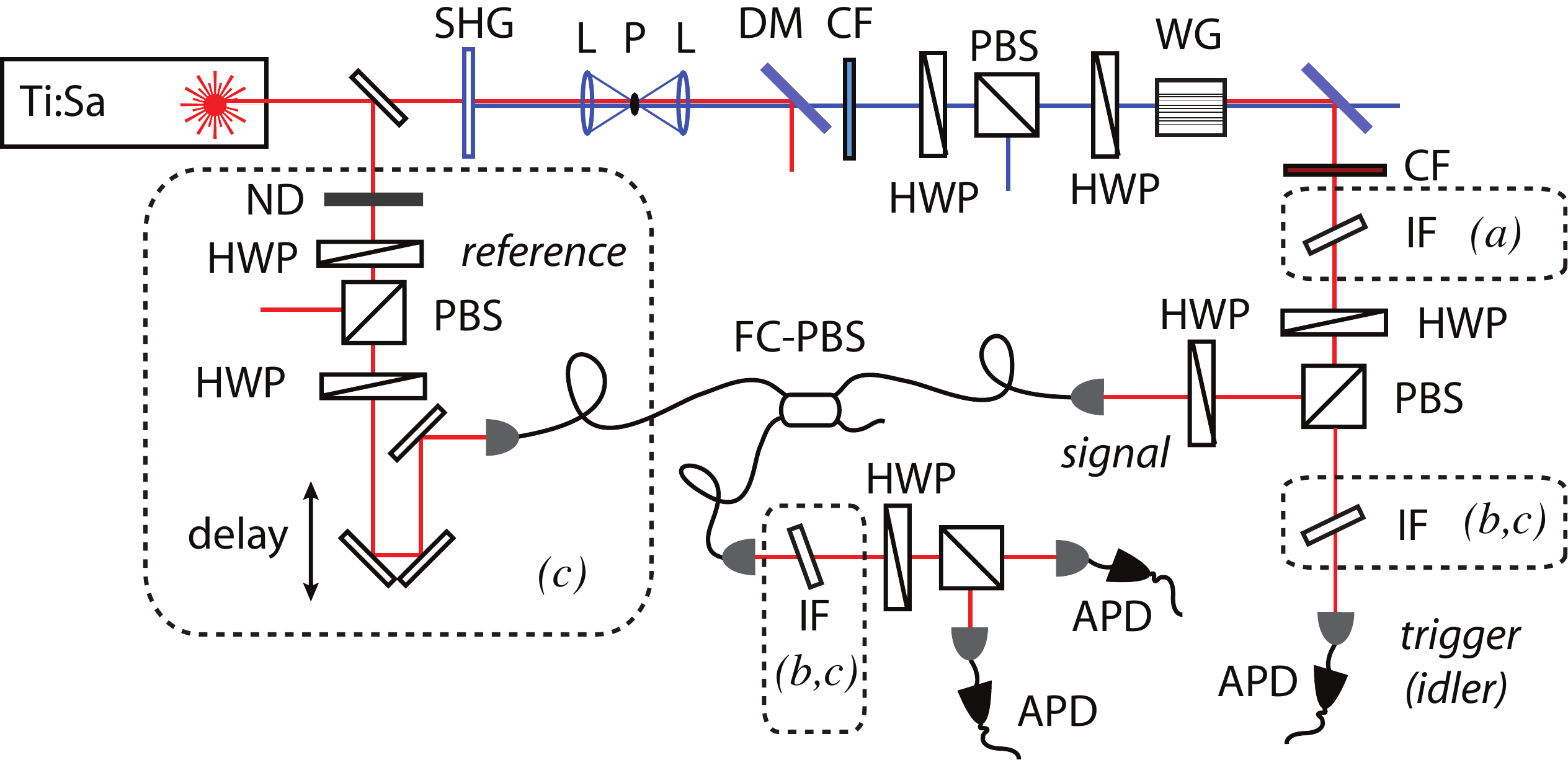}
\caption{Setup for preparing and characterizing single photons from waveguided PDC source applied for (a) measuring HOM interference between signal and idler, (b) recording the statistics of the heralded state, and (c) measuring the HOM interference between two independent sources. Abbreviations: SHG = second harmonic generation (0.5 mm thick BBO),  L = lens, P = pinhole,  DM = dicroich mirror,  CF = color glass filter,  HWP = half-wave plate, PBS = polarizing beam splitter, WG = waveguide,   IF = interference filter, ND = neutral density filter, FC-PBS = fiber coupled PBS, APD = avalanche photo diode, SG = spectrograph$^{*}$  and SM = spectrometer$^{*}$.   ($^{*}$ only shown in Figs~\ref{fig_marginals} and \ref{fig_shg_response}.)
}
\label{fig_setups}
\end{center}
\end{figure}

Since the twin-photons  of a type-II process have orthogonal polarizations, we measured HOM interference fringes between signal and idler by rotating a half-wave plate in front of a polarizing beam splitter and recording the coincidence counts between the two output arms. The experiment was realized in three different cases: (I) without spectral filter, (II) including in the setup a spectral filter with FWHM of \unit{2.5}{\nano\meter} and, (III) decreasing the width of the filter down to   \unit{1}{\nano\meter}. The filters were inserted in the joint beam path of signal and idler as shown in Fig.~\ref{fig_setups}(a). According to auxiliary experimental observations, the time delay between p- and s-polarized modes is on the order of \unit{700}{\femto\second}. Since the temporal width of signal and idler, estimated by the  filter bandwidth, is on the order of \unit{1-2}{\pico \second} and no delay line is inserted in the setup, the temporal overlap cannot be neglected in our measurements.

\begin{figure}[!htb]
\begin{center}
\includegraphics[width = 0.5\textwidth]{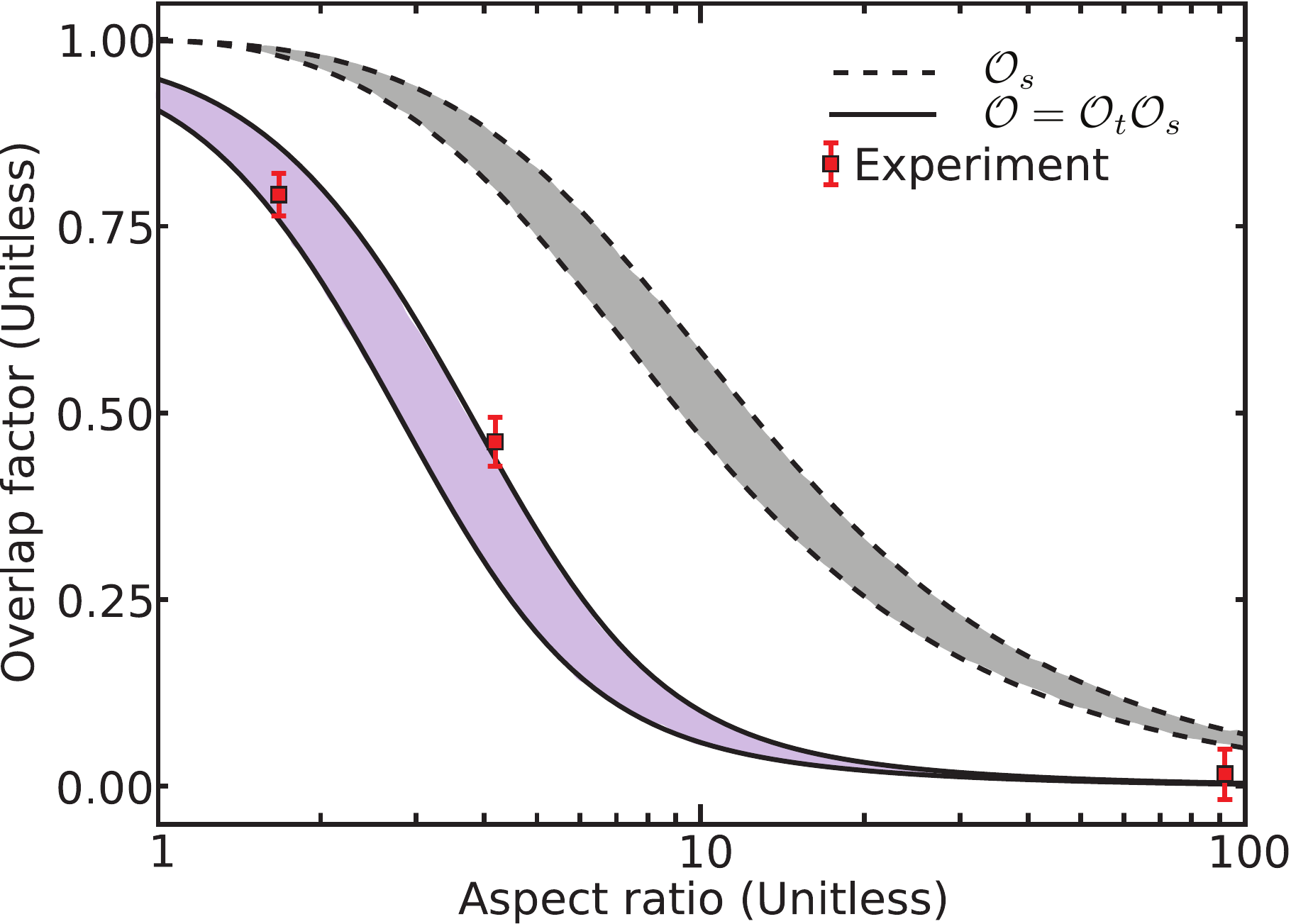}
\caption{Measured overlap values (squares) together with theoretical estimation for the spectral (dashed line) and total (solid line) overlaps with respect to the aspect ratio. We consider the PM tilt of $\theta = 54.7^{\circ}\pm1.5^{\circ}$ (indicated with the shaded areas).}
\label{fig_aspect}
\end{center}
\end{figure}

The visibility is defined by $V = ({C}_{max}-{C}_{min})/({C}_{max}+{C}_{min})$, where $C$ indicates the amount of coincident counts. In terms of the overlap factor, the visibility can also be expressed as $V = (1+\mathcal{O})/(3-\mathcal{O})$. The classical case, in which signal and idler are completely distinguishable,  corresponds to the situation of  zero overlap, i.e. $V=1/3$.
We measured, without any background subtraction, the following visibilities: $V(I) = 0.34 \pm 0.02$,  $V(II)= 0.58 \pm 0.02$,  and $V(III)= 0.81 \pm 0.03$, corresponding to each of the above mentioned cases. The corresponding overlap factors must be compared with the theoretical prediction given in Eqs~(\ref{eq_O_spectral}) and (\ref{eq_O_temporal}). For this we approximate the aspect ratios in a straightforward manner. When no filtering is applied, $\mathcal{A}$ is estimated as the ratio of the experimentally extracted major and minor axes, i.e, $\mathcal{A}(I) \sim 90-100$. When spectral filtering is applied, we determine $\mathcal{A}$ as the ratio of the filter bandwidth to the PM width i.e.,  $\mathcal{A}(II) \sim 4.2$ and $\mathcal{A}(III) \sim 1.7$. As expected,  the spectral overlap alone does not explain our results, as can be seen in Fig.~\ref{fig_aspect}.  By including the temporal overlap in the analysis, the employed model predicts  the right tendency, even though in the experiment we do not have any compensation for dispersion or chirp.

We conclude that our model, which includes only experimentally extracted characteristics of the PM, explains with high confidence results of an independent experiment. In addition, it predicts how the spectral correlation between signal and idler can be manipulated with the selected filter.  
In the next sections we study how the heralded state is affected by the higher photon-number contributions and remaining spectral correlation.

\section{\label{sec_statistics}Measurement of the heralded statistics}
In the low gain regime heralding of  pure signal states  is normally realized by considering a single spectral filter at the trigger.
However, at higher gains multiple modes are excited simultaneously. Thus,  filtering only the  trigger results in the detection of uncorrelated signal modes. 
In addition to this, the higher photon number contributions decrease the fidelity of the heralded single-photon state. We studied these effects by measuring, how the heralded statistics depends on whether or not a narrowband spectral filter is added in the signal arm.
As before, we pumped our PDC source with a frequency doubled Ti:Sapphire laser. The pump beam was sent through a simple spatial mode cleaner before being launched  into the WG. After that, the pump was filtered out and the twins were divided into two independent paths. The idler was sent through \unit{1}{\nano\meter} broad  interference filter and, finally, coupled into SM fiber.

We measured the heralded statistics in two different configurations: either without filter or with a 2.5~nm interference filter inserted in the signal arm.  The setup shown in Fig. \ref{fig_setups}(b) was arranged such  that the idler was used as a trigger and the statistics of the signal was measured with a ``two-bin TMD''. This detector consists of a single 50/50 beam splitter, constructed by using a half-wave plate in front of a polarizing beam splitter. The loss tolerance of the TMD was achieved by accurate calibration of loss $\boldmath{L}(\eta)$ and convolution  matrices $\boldmath{C}$ \cite{D.Achilles2004}. The signal statistics $\vec{\varrho}$ was then deduced by inverting the click statistics $\vec{p}_{click}  = \boldmath{CL(\eta)}\vec{\varrho} $ with a maximum-likekihood technique \cite{Banaszek1998}.

\begin{figure}
\begin{center}
\includegraphics[width = 1.0\textwidth]{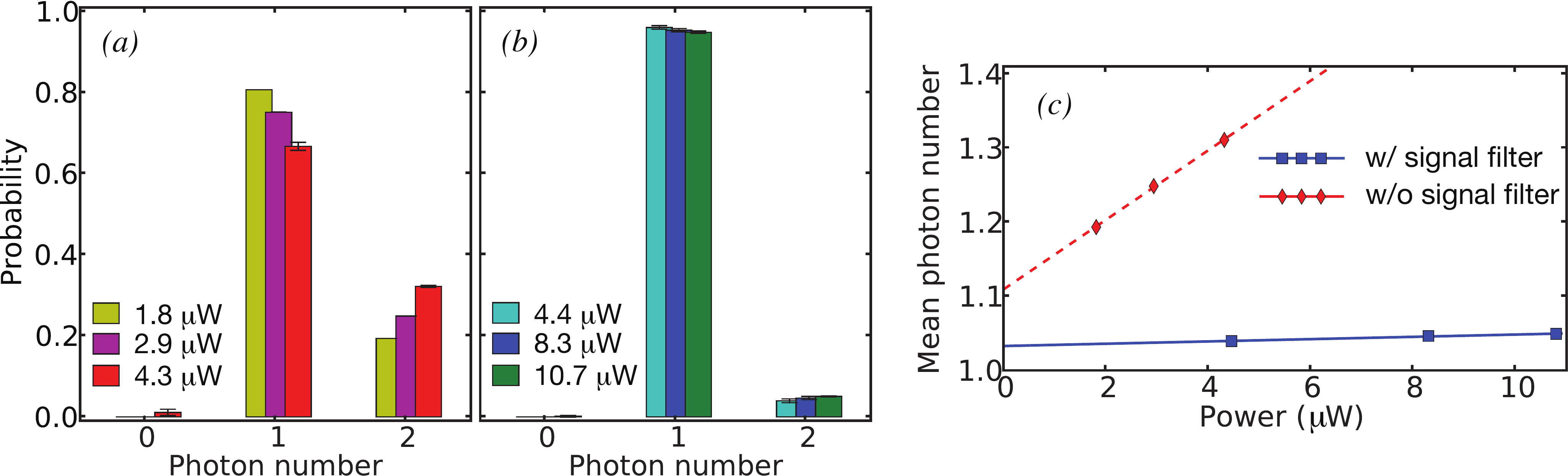}
\caption{Heralded statistics  at different pump powers with  (a) no filter and (b) \unit{2.5}{\nano\meter} filter in signal. (c) Mean photon number of the statistics with respect to the pump power: diamonds referring to plot (a) and squares to plot (b) are shown together with linear fits to the data.}
\label{fig_statistics_filtered}
\end{center}
\end{figure}

From our results one can observe that the heralded statistics presents a large two-photon contribution [Fig.~\ref{fig_statistics_filtered}(a)], which is reduced by placing a narrow band filter in the signal arm [Fig.~\ref{fig_statistics_filtered}(b)]. We understand this result as an indication that, at higher gains, a large amount of distinct spectral modes are excited simultaneously. If two idler photons are generated there is a high probability that just one of them  has a good overlap with the filter mode. Therefore, by using a filter in the signal arm it is possible to reject the signal photon created together with the spectrally mismatched idler. The click statistics were inverted using the calibrated efficiencies, which in both cases were $9\%$, corresponding to the ratio of coincident counts to singles. The highest trigger rate was \unit{5.9}{\kilo \hertz}, which is equivalent to a heralding probability of 0.16\%. Regarding the single-photon state preparation,  the one-photon contribution of the filtered signal was 95\% at this trigger rate.

A comparison between the filtered and unfiltered statistics allows us to extract the degree of mode reduction given by the filter. For that, we model the multimode statistics as the convolution of several uniformly occupied spectral modes~\cite{W.P.Grice2001, A.M.Branczyk2009}. We begin the analysis with the single-mode case. The heralded statistics, being conditioned on a trigger event, has the form
\begin{equation}
P_{n | click} = \frac{P_{n \cap  click}}{P_{click}}\;.
\end{equation}
Moreover, the  joint probability of  having $n$ photons in signal and detecting a click from idler is given by $P_{n  \cap click} = P^{PDC}_{n}[1-(1-\eta_{t})^{n}]$, where $\eta_{t}$ is the efficiency at which the filtered mode is detected and $P^{PDC}_{n}$ the probability of creating $n$ signal (idler) photons, which always occur in pairs for PDC.  Since in the experimental realization losses are very high, the joint probability can be approximated  by $P_{n  \cap click} \approx  \eta_{t} n   P^{PDC}_{n} $. The  overall click probability is estimated as $P_{click} = \sum_{n} P_{n  \cap click} \approx \eta_{t} \left <n \right >$, where $\left <n \right > $ indicates the mean photon number of the marginal PDC statistics. In this loss regime, the heralded statistics become independent of the trigger efficiency $\eta_{t}$ and its mean photon number can be evaluated as
$\left <n_{\mathrm{heralded}} \right > = \sum_{n} n P_{n|click}  = \left <n^{2} \right > / \left <n \right >$, with $\left <n^{2} \right > $ being the second moment of the marginal PDC statistics.

In the case of a multimode PDC with several uniformly excited modes, the statistics $ P^{PDC}_{n}$ is given by the convolution of the modes, each having the mean photon number $ \left <n_{sm} \right > $ and the second moment  $\left <n_{sm}^{2} \right > $. For $M$  excited modes we find that  $ \left <n \right > = M \left <n_{sm} \right >  $ and $ \left <n^{2}\right > = M \left <n^{2}_{sm} \right > +M(M-1)  \left <n_{sm} \right >^{2} $, from which we can calculate the multimode mean photon number of the heralded statistics. Considering a thermal single mode marginal distribution,  $\varrho^{sm}_{n} = |\chi|^{2n}(1-|\chi|^{2})$ with $|\chi|$ being the gain of the PDC mode, 
the mean photon number of the heralded statistics can be approximated by
\begin{eqnarray}
\left <n_{\textrm{heralded}} \right > &=&  \frac{\left <n^{2}_{sm} \right >}{\left <n_{sm} \right >} +(M-1)\left <n_{sm} \right > \nonumber \\
& \approx & 1+(M+1)|\chi|^{2}\;.
\end{eqnarray}
We conclude that, in the limit of low pump powers,  such that $|\chi|^{2} \propto \textrm{pump power}$,
the slope of the mean photon number depends on the number of modes.  In  Fig.~\ref{fig_statistics_filtered}(c) we show the mean photon number of the measured heralded statistics with respect to the pump power. By comparing the slopes, we estimate that the reduction in the number of modes due to the filter at signal is on the order of 30.
We emphasize that the model used is an approximation, where all modes are treated similarly, even though the trigger mode is a special one.

\section{\label{sec_HOM}Interference between  independent sources}
In this section we study the HOM interference between two independent sources: the heralded signal and a strongly attenuated coherent field, here called reference. The visibility of the interference dip depends, on the one hand, on the presence of higher photon-number components and, on the other hand, on the spectral overlap between the target state and the reference. 
We experimentally investigate the cases of two- and three-fold HOM interferences. The former case corresponds to the interference between signal and reference neglecting the PDC photon-number correlation, while in the latter case  the signal state is heralded by trigger events. 

We begin our study by considering a  spectrally pure signal state. Further, we decompose the signal only up to the second order in the photon-number basis, which is justified under the experimental conditions. After that, we complete our description by adding the effect of spectral correlations between signal and idler. In  the experiment, signal and reference are launched into the ports $a$ and $b$ of a 50/50 beam splitter. The density matrix of the input state $\hat{\rho}_{ab}$ can be written as
\begin{equation}
\hat{\rho}_{ab} = \Big ( p_{0}\ket{0}  \bra{0}+ p_{1}\ket{1} \bra{1} +p_{2}\ket{2}  \bra{2}  \Big)_a \otimes \Big( \ket{\beta} \bra{\beta} \Big)_b\,,
\label{eq_photon_state}
\end{equation}
in which $p_{n}$ ($n = 0,1,2$)  corresponds to the statistics of the photon-number mixed signal mode with  $\sum_{n} p_{n} = 1$. The reference field has amplitude $\beta$ and spectral distribution $u(\omega)$, such that $\int \hspace{-2pt} d\omega \,|\beta \,u(\omega)|^{2}= |\beta|^{2}$. Since 
mode ``$a$'' is spectrally pure, its  one- and two-photon components can be respectively described by $\ket{1}\,\subr{a} = \int \hspace{-2pt} d\omega \, f\hspace{-1pt}(\omega)\, \hat{a}^{\dagger}\hspace{-1pt}(\omega)\ket{0}\,\subr{a}$ and $\ket{2}\,\subr{a} = \frac{1}{\sqrt{2}}\int \hspace{-2pt} d\omega \int \hspace{-2pt} d \tilde{\omega} \, f\hspace{-1pt}(\omega) f\hspace{-1pt}(\tilde{\omega}) \, \hat{a}^{\dagger}\hspace{-1pt}(\omega)\, \hat{a}^{\dagger}\hspace{-1pt}(\tilde{\omega})\ket{0}\,\subr{a}$, in which the spectral distribution $f \hspace{-1pt}(\omega)$ is normalized, i.e. $\int \hspace{-2pt} d\omega \, |f \hspace{-1pt}(\omega)|^{2}= 1$.

Interference does not take place when the two input pulses arrive at the beam splitter at different times. To model that, we explicitly write the  evolution of the fields considering a time delay $\tau$. In the frequency domain, the beam splitter relations are given by
\begin{eqnarray}
\hat{a}^{\dagger}(\omega)  &=&\frac{1}{\sqrt{2}} \left [\hat{c}^{\dagger}(\omega)-\hat{d}^{\dagger}(\omega)\right]e^{i\omega \tau} ,\nonumber \\
\hat{b}^{\dagger}(\omega) &= &\frac{1}{\sqrt{2}}\left[ \hat{c}^{\dagger}(\omega)+\hat{d}^{\dagger}(\omega)\right]\,,
\end{eqnarray}
in which $c$ and $d$ are the two output modes.  These equations allow us to calculate the evolved state $\hat{\rho}_{cd}$. Finally, the probability of a coincident click can be calculated via
$P = Tr \left \{ \hat{\rho}_{cd}\  \hat{\Pi}_{c}\otimes \hat{\Pi}_{d}\right\} $, where $ \hat{\Pi}_{c}\otimes \hat{\Pi}_{d} = (\mathds{1} - \ket{0}\,\subs{c}{c}\,\bra{0})\otimes (\mathds{1} - \ket{0}\,\subs{d}{d} \,\bra{0}) $ describes the POVM of two simultaneous clicks in the APDs. 
Using the bosonic commutation relation $[\hat{a}(\omega), \hat{a}^{\dagger}(\tilde{\omega})]= \delta(\omega-\tilde{\omega})$ and the fact that $\hat{b}(\omega)\ket{\beta} = \beta u({\omega})\ket{\beta}$, the probability of detecting a coincident click is given by 
\begin{eqnarray}
P&=& p_{0}(1 -e^{-|\beta^{\prime}|^{2}})^{2} +p_{1}(1-e^{-|\beta^{\prime}|^{2}}) - p_{1}\,T(\tau)\,|\beta^{\prime}|^{2}\,e^{-|\beta^{\prime}|^{2}}\nonumber \\
 &\phantom{=} &  +\;p_{2}(1-\frac{1}{2}e^{-|\beta^{\prime}|^{2}}) - p_{2}\,e^{-|\beta^{\prime}|^{2}}\, (\;T(\tau) |\beta^{\prime}|^{2}+ T^{\prime}(\tau) \,\frac{|\beta^{\prime}|^{4}}{4}\;)\,,
 \label{eq_P}
\end{eqnarray}
with $\beta^{\prime} = \beta / \sqrt{2}$. The functions $T(\tau)$ and $T^{\prime}(\tau)$ have the form
\begin{eqnarray}
\label{eq_overlapfacators}
T(\tau) \hspace{-7pt}&=&\hspace{-8pt}\int \hspace{-2pt}d \omega_{1} \hspace{-2pt} \int \hspace{-2pt}d \omega_{2} \ u^{*}(\omega_{1}) \ g(\omega_{1}, \omega_{2})  \ u(\omega_{2})e^{i \tau( \omega_{1}-\omega_{2})} \,, \\
T^{\prime}(\tau)\hspace{-7pt}&=&\hspace{-8pt}\int \hspace{-2pt}d \omega_{1}  \hspace{-2pt}  \int \hspace{-2pt}d \omega_{2} \hspace{-2pt}   \int \hspace{-2pt}d\omega_{3}  \hspace{-2pt}  \int \hspace{-2pt}d \omega_{4} \
u^{*}(\omega_{1})u^{*}(\omega_{2}) \  h(\omega_{1},\omega_{2},\omega_{3}, \omega_{4})  \ u(\omega_{3})u(\omega_{4})e^{i \tau( \omega_{1} + \omega_{2} - \omega_{3} - \omega_{4})}, \nonumber 
\end{eqnarray}
where $g(\omega_{1}, \omega_{2}) = f(\omega_{1})f^{*}(\omega_{2})$ and 
$ h(\omega_{1},\omega_{2},\omega_{3}, \omega_{4}) =  f(\omega_{1}) f(\omega_{2}) f^{*}(\omega_{3})  f^{*}(\omega_{4}) $.
At origin ($\tau = 0$) these factors determine the spectral overlap  of the one- and two-photon wave-packets with respect to the reference.
Finally, the visibility of the interference dip is defined by 
\begin{equation}
\mathcal{V}= \frac{P(\tau\rightarrow \infty) -P(\tau = 0)}{P(\tau\rightarrow \infty) }\,.
\label{eq_visibility_def}
\end{equation}

\begin{figure}
\begin{center}
 \includegraphics[width = 0.75\textwidth]{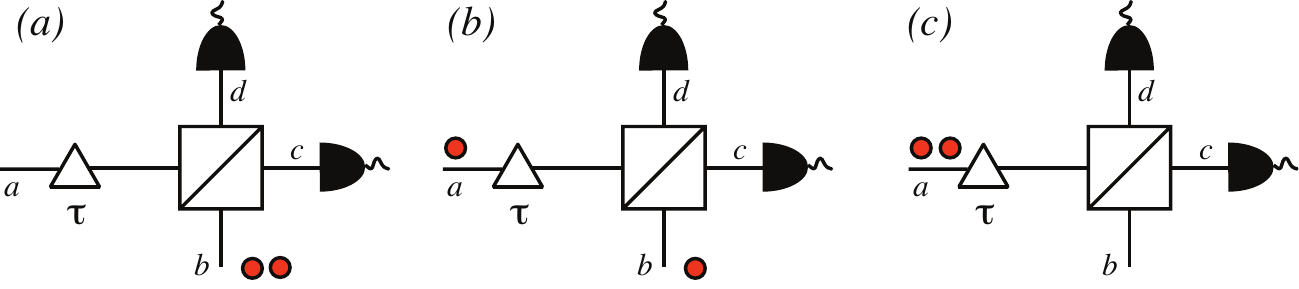}
\caption{Contributions affecting the HOM interference between signal and reference.}
\label{fig_ball_model}
\end{center}
\end{figure}

In order to simplify our model we compare the independent impact of the one- and two-photon components of signal on the visibility. Considering perfect spectral overlap between these components and the reference field, and appliying values of $|\beta|^{2}$ compatible with the experimental realization, we estimate that the interference of the one-photon component  would have a visibility of above 98\%, whereas the one of the two-photon component would be less than 5\%. Therefore, in the regime of power considered here, the two-photon component of signal gives only
a background contribution to the coincidences. Thus, we consider the spectral degree of freedom exclusively for the one-photon component of signal and reference. Further, we approximate $e^{-|\beta^{\prime}|^{2}} \approx 1-|\beta^{\prime}|^{2}$, which is well valid within the experimental frame, and neglect the terms arriving from the interference of more than two photons (total number in mode ``$a$'' plus ``$b$''). The remaining contributions are shown in Fig.~\ref{fig_ball_model} and they can be retrieved from Eq.~(\ref{eq_P}) as follows.  The first term corresponds to accidental counts due to the two-photon component of the reference and it can be re-expressed as $p_{0}(1-e^{-|\beta^{\prime}|^{2}})^{2} \approx p_{0}|\beta^{\prime}|^{4}$ [Fig.~\ref{fig_ball_model}(a)]. The second and third terms account for the events where both signal and reference contain one single photon [Fig.~\ref{fig_ball_model}(b)]. 
The probability of simultaneous accidental clicks due to one photon from signal and another from reference is given in the second term as $p_{1}(1-e^{-|\beta^{\prime}|^{2}}) \approx p_{1}|\beta^{\prime}|^{2}$. Considering the first-order photon-number contribution, the third term describing the quantum interference between the one-photon components of signal and reference can be rewritten as
$p_{1}T(\tau)|\beta^{\prime}|^{2}$. The fourth term takes  into account  accidental counts coming from signal, 
$p_{2}(1-\frac{1}{2}e^{-|\beta^{\prime}|^{2}}) \approx \frac{1}{2}p_{2}$ [Fig.~\ref{fig_ball_model}(c)]. The last term, corresponding to  interference of at least three photons, is an order of magnitude smaller the other terms and thence it is neglected.

With the described simplifications, Eq.~(\ref{eq_P}) can be rewritten as
\begin{equation}
P(\tau) = p_{0}|\beta^{\prime}|^{4} + p_{1}|\beta^{\prime}|^{2}(1-T(\tau)) +\frac{1}{2}p_{2}\,.
\label{eq_PT}
\end{equation}
After substituting Eq.~(\ref{eq_PT}) into Eq.~(\ref{eq_visibility_def}), the visibility can be written with respect to the mean photon number of the reference $|\beta|^{2}$ as
\begin{equation}
\mathcal{V}(|\beta|^{2})= \frac{p_{1}T(\tau= 0)}{p_{0}\frac{|\beta|^{2}}{2} +p_{1}+ p_{2}\frac{1}{|\beta|^{2}}},\,
\label{eq_visibility}
\end{equation}
which reaches its maximum at $|\beta|^{2}_{\textrm{max}} = \sqrt{2\,p_{2}/p_{0}} $.

In the experimental situation the joint-spectral distribution of signal and idler is not separable, and thus the prepared state is spectrally mixed. Our model can be adapted to this condition by replacing the function $g(\omega_{1}, \omega_{2})$ by a new one, in which a trace operation over the idler mode is realized. The spectral density of the PDC state after filtering is  given by
\begin{equation}
 g_{\mathrm{\,pdc}}(\omega_{1}, \omega_{2}) = \frac{1}{N} \,\left( \int d\omega_{i} \ \phi(\omega_{1}, \omega_{i}) \phi^{*}(\omega_{2}, \omega_{i})t_{i}(\omega_{i}) \sqrt{t_{s}(\omega_{1})}\sqrt{ t_{s}(\omega_{2})}\, \right),
 \label{eq_g_PDC}
 \end{equation}
where the $t_{\mu}(\omega)$ is the intensity transmission profile of the signal and idler filters employed in the state generation and $N$ accounts for the normalization. 

Apart from the visibility, the width of the HOM interference dip offers information about the spectral properties of the prepared state. Using the Gaussian approximation for the spectral correlation function, reference, and filter profiles, the width of the HOM interference dip can be estimated via Eq.~(\ref{eq_PT}). Disregarding the background contributions, it can be written as
\begin{equation}
P_{\textrm{spectral}}(\tau) \propto 1- T(\tau) =  1-\,T^{\textrm{max}}\, \exp \hspace{-3pt} \left( -\frac{\tau^{2}}{2 \,\sigma^{2}_{t}}\right)\,,
\label{eq_tau}
\end{equation}
where the temporal width of the interference dip, $\sigma^{2}_{t}= \frac{1}{\sigma^{2}}+ \frac{1}{\sigma^{2}_{\beta}}+\frac{1}{\sigma^{2}_{t_{s}}}+\frac{\sin^{2}{\theta}}{\sigma^{2}_{pm}}$, is determined by the inverse widths of the pump $\sigma$, reference $\sigma_{\beta}$, signal filter $\sigma_{t_{s}}$, and  phase matching $\sigma_{pm}$, as well as by the ellipse's tilt $\theta$. The amplitude $T^{\textrm{max}}$ describes the maximal expected value of the overlap, ranging from zero to one.  We emphasize that the two- and three-fold interference dips have the same temporal width.

For the measurement of the HOM interference  we used the setup shown in Fig.~\ref{fig_setups}(c). A portion of the original laser beam was heavily attenuated and launched  through a polarization maintaining SM fiber-coupler in s-polarization. The signal was launched to the fiber in p-polarization and overlapped with the reference at  a 50/50 beam splitter, consisting of a half-wave plate and a polarizing beam splitter.  Two 1~nm broad  interference filters  were inserted in the setup: one in the path of the trigger and the other in the joint arm of the signal and reference. We measured the HOM interference dip in the two- and three-fold cases for different amplitudes of the reference. By blocking the reference arm we measured the statistics of the signal state and obtained the results  summarized in Table \ref{table_statistics}.  

\begin{table}
\begin{center}
\begin{tabular}{cccc}
\toprule
 &$p_{0}$(\%)&$p_{1}$(\%)&$p_{2}$(\%)\\
\midrule
2-fold &99.7896 &0.2101 & 0.0003  \\
3-fold & 94.920& 5.065 & 0.015 \\
\bottomrule
\end{tabular}
\caption{Measured signal state statistics $p_{n}$ $ (n= 0,1,2)$ without loss inversion.}
\label{table_statistics}
\end{center}
\end{table}

\begin{figure}[!htb]
\begin{center}
 \includegraphics[width = 0.5\textwidth]{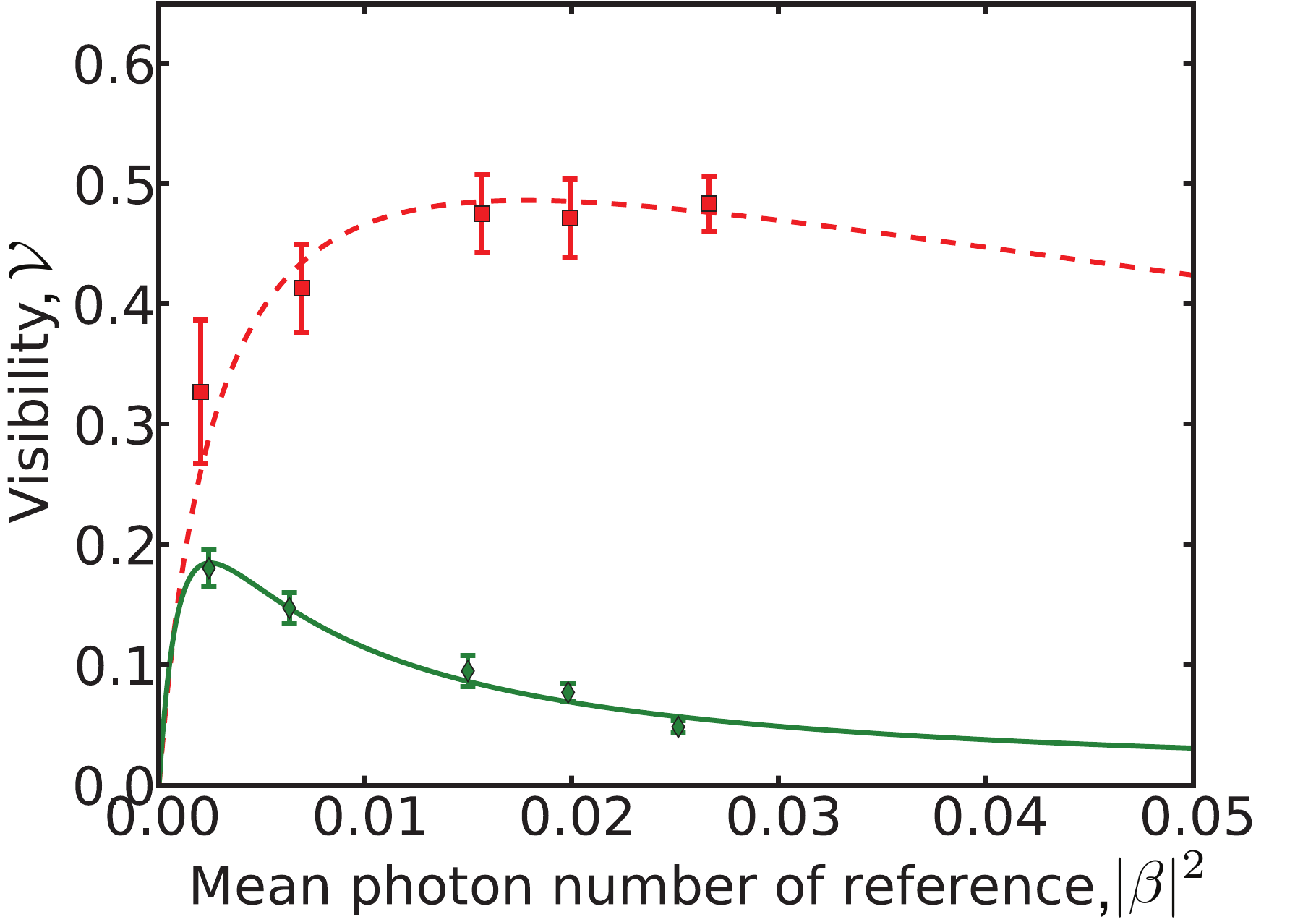}
\caption{Visibility of the HOM interference dip with respect to the mean photon number of reference.
Measured points are marked by diamonds  (two-fold interference) and squares (three-fold interference). Solid and dashed lines correspond to the fitted curves from which the spectral overlap value can be extracted. }
\label{fig_overlap}
\end{center}
\end{figure}

Substituting the measured statistics back into Eq.~(\ref{eq_visibility})  we can find the maximal expected visibilities of $\mathcal{V}^{\textrm{max}}_{\textrm{2-fold}}= 0.46$ and $\mathcal{V}^{\textrm{max}}_{\textrm{3-fold}}= 0.75$. In these values we disregard the spectral mismatch, i.e. $T = 1$, and consider only the degradation due to the two-photon component.  Our experimental results are shown in Fig.~\ref{fig_overlap}, where we plot the measured visibilities with respect to the mean photon number of the reference beam $|\beta|^{2}$. The latter can be found by measuring the single-counts rate of the reference in one of the detectors while blocking the signal,  $P_{\mathrm{singles}} = 1 - e^{-1/2 |\beta|^{2}}\approx |\beta|^{2}/2$. The discrepancy between the expected and observed visibilities is a consequence of the spectral mismatch. We extract the value of the spectral overlap $T$ from the measurement by fitting Eq.~(\ref{eq_visibility}) to the experimental data. The obtained overlap values are $T_{\textrm{2-fold}}= 0.41\pm 0.02$ and $T_{\textrm{3-fold}}= 0.65\pm{0.03}$. This clearly shows the increase in purity achieved via filtered herald, i.e  the degree of  spectral correlation between the twins is reduced for the selected conditioned events. Finally, the fidelity of the prepared state is given by $\mathcal {F} = \sqrt{\braket{1| \hat{\rho}|1}} =  \sqrt{T\varrho_{1}}$, where $T$ is the spectral overlap and $\varrho_{1}$ is the one-photon component of the inverted statistics shown in Table \ref{table_inverted}. Combining the results of both measurements we find that  the fidelity of the heralded state was $\mathcal{F}_{\textrm{3-fold}} = 0.78\pm0.03$.

\begin{table}
\begin{center}
\begin{tabular}{cccc}
\toprule
 &$\varrho_{0}$(\%) &$\varrho_{1}$(\%) &$\varrho_{2}$(\%) \\
\midrule
3-fold & 0.2$\pm0.2$ & 93.1$\pm0.6$ & 6.6$\pm0.6$\\
\bottomrule
\end{tabular}
\caption{Statistics $\varrho_{n}$ of the heralded state in Table \ref{table_statistics} inverted with $4.8\%\pm 0.2\%$ efficiency.}
\label{table_inverted}
\end{center}
\end{table}

\begin{figure}[!htb]
 \begin{center}
\includegraphics[width = 1.0\textwidth]{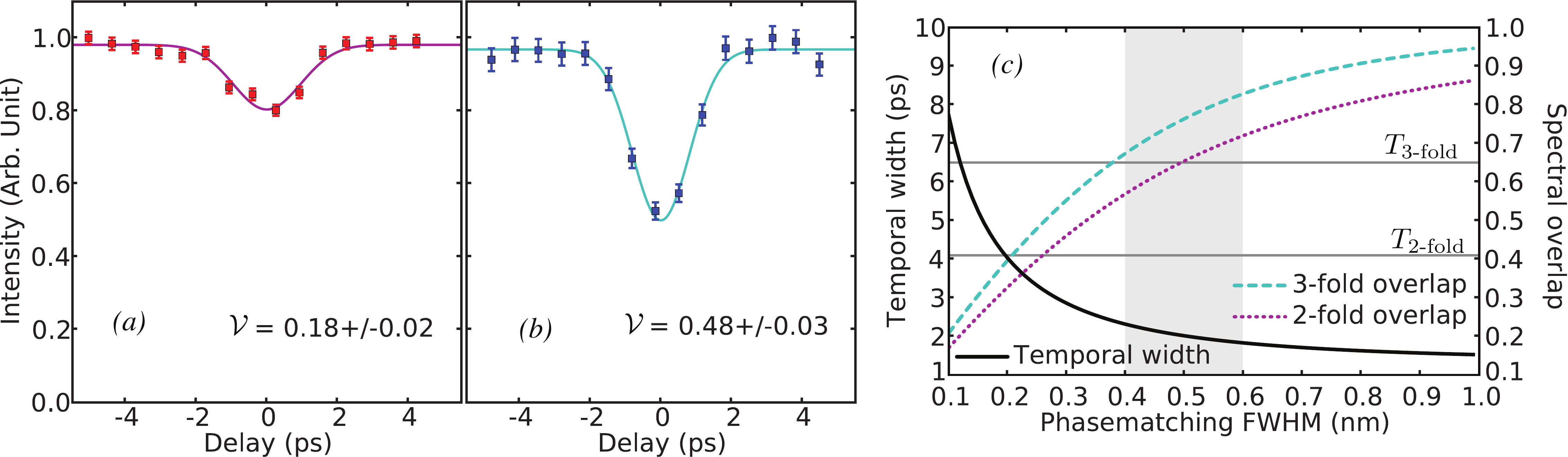}
\caption{Measured HOM interference dips between the prepared signal and the reference in (a) two-  and (b) three-fold  cases. (c) Theoretical estimation of the temporal width (black solid line) and maximal spectral overlaps (three-fold: dashed line, two-fold: dotted line) with respect to the PM bandwidth (employed FWHM bandwidths: pump \unit{2.5}{\nano\meter}, reference and filters \unit{1}{\nano\meter}, and PM tilt: $\theta = 54.7^{\circ}).$ Horizontal lines refer to the measured values of overlap. Shaded area indicates the FWHM of the PM estimated by  the temporal dip widths.
}
\label{fig_3-fold}
\end{center}
\end{figure}

In Figs~\ref{fig_3-fold}(a) and \ref{fig_3-fold}(b) we show example measurements of the two- and three-fold interference dips that were recorded with reference powers close to the optimized values $|\beta|^{2}_{\textrm{max}}$  in Fig.~\ref{fig_overlap}. The measured visibilities were $\mathcal{V}_{\textrm{2-fold}} = 0.18\pm0.02$ and $\mathcal{V}_{\textrm{3-fold}} = 0.48\pm0.03$, respectively. The temporal width of the interference dips  was $2.0\pm 0.2$~ps. 
Following Eq.~(\ref{eq_tau}) we theoretically study how this width changes as a function of the PM bandwidth [Fig.~\ref{fig_3-fold}(c)]. The comparison of the measurement with our model suggests a FWHM of \unit{0.5\pm0.1}{\nano\meter} for the PM, which is in good agreement with the value extracted from SH measurements. Further exploring our knowledge about the spectral correlation function, we use  Eqs.~(\ref{eq_overlapfacators}) and (\ref{eq_g_PDC}) in order to estimate the maximal spectral overlap $T^{\textrm{max}}$ in the two- and three-fold cases. In Fig.~\ref{fig_3-fold}(c) we illustrate the theoretical predictions
and  compare them with the measured overlap values. In the three-fold case we find a good agreement with the  model.  However, the measured value of the two-fold overlap deviates largely from the expected one. As no selection of the detection events by triggering is applied in the two-fold case, we believe that the properties of non-ideal WG, such as  fluorescence and higher order spatial modes with complex spectral structure, degrade the result. In addition to this, the expected values have not been achieved due to other experimental imperfections,  in particular mechanical instabilities and spatial mode mismatch. 

\section{Conclusions}
We have studied the ability to tailor the spectral properties of a conventional  PP-KTP waveguide. Information about the spectral correlation between signal and idler was retrieved from  simple experiments, enabling to model our system with a standard theory of parametric down-conversion. The viability of this model was tested via the Hong-Ou-Mandel interference between the twin beams.
Since the production of more evolved non-Gaussian states, as displaced single-photon Fock states, 
requires  an overlap between the signal and an independent reference field, we  studied the quantum interference between the heralded  state and a coherent reference field.  The spectral overlap was directly extracted  from the visibility of the Hong-Ou-Mandel interference.
 Further, we experimentally characterized the statistics of the  heralded state. 
  Considering both the spectral overlap and the one-photon contribution,  we obtained by a direct detection the fidelity of 78\% between the prepared state and a single-photon Fock state. In addition, our observations show that heralding decreases the number of modes only if both signal and idler are spectrally filtered.  
Although we found that in the current experiment  the spectral overlap between the reference and the signal state is not yet optimized, we provide a clear and simple recipe for tailoring the characteristics of the spectral correlation  and for testing the degree of decoupling, i.e., the fidelity of the heralded state.

\section*{Acknowledgments}
We thank M.~Avenhaus,  A.~Christ  and P.~J.~Mosley for valuable and helpful discussions. This work was supported by the EC under the grant 
agreement CORNER (FP7-ICT-213681). K.N.C.~acknowledges the support from the Alexander von Humboldt Foundation.


\vspace{3mm}

\end{document}